\documentclass[12pt]{iopart}

\makeatletter
\def\convertto#1#2{\strip@pt\dimexpr #2*65536/\number\dimexpr 1#1}
\makeatother

\makeatletter
\newcommand\thefontsize[1]{{#1 The current font size is: \f@size pt\par}}
\makeatother

\usepackage{graphicx}
\usepackage{citesort}
\usepackage{cancel}
\newcommand{\binom}[2]{{#1 \choose #2}}
\usepackage{framed,color}

\begin{document}

\title[Preparation of mesoscopic atomic ensembles with single-particle resolution]{Preparation of mesoscopic atomic ensembles with single-particle resolution}

\author{A. H\"uper$^1$, C. P\"ur$^1$, M. Hetzel$^1$, J. Geng$^1$, J. Peise$^1$, I. Kruse$^1$, M. Kristensen$^2$, W. Ertmer$^1$, J. Arlt$^2$ and C. Klempt$^1$}

\address{$^1$ Institut f\"ur Quantenoptik, Leibniz Universit\"at Hannover, Welfengarten 1,\\30167 Hannover, Germany\\
$^2$ Department of Physics and Astronomy, Aarhus University, 8000 Aarhus C, Denmark}
\ead{hueper@iqo.uni-hannover.de}
\vspace{10pt}
\begin{indented}
\item[]December 11th, 2019
\end{indented}

\begin{abstract}
The analysis of entangled atomic ensembles and their application for interferometry beyond the standard quantum limit requires an accurate determination of the number of atoms.
We present an accurate fluorescence detection technique for atoms that is fully integrated into an experimental apparatus for the production of many-particle entangled quantum states.
Single-particle resolving fluorescence measurements for $1$ up to $30$ atoms are presented.
According to our noise analysis, we extrapolate that the single-atom resolution extends to a limiting atom number of $390(20)$ atoms.
We utilize the accurate atom number detection for a number stabilization of the laser-cooled atomic ensemble.
For a target ensemble size of $7$ atoms prepared on demand, we achieve a $92(2)\,\%$ preparation fidelity and reach number fluctuations $18(1)\,\mathrm{dB}$ below the shot noise level using real-time feedback on the magneto-optical trap.
\end{abstract}

%
%
%
%
%

\section{Introduction}
Large systems of entangled particles can be built by adding more and more constituents and by engineering the entanglement between them.
Alternatively, large numbers of up to 3,000 mutually entangled ultracold atoms~\cite{McConnell2015,Haas2014,Lucke2014,Hosten2016} can be generated by exploiting the indistinguishability of the atoms.
To harness the full potential of such systems, the conceptual and technological challenge is to control the number of indistinguishable atoms on the single-particle level.

A prime example is the application of entangled atomic ensembles for atom interferometry beyond the standard quantum limit (SQL)~\cite{Pezze2018}.
Atom interferometers allow measuring a quantity of interest (e.g. electromagnetic fields, time, acceleration, rotation, gravitational fields) by detecting the relative phase $\theta$ between two atomic states.
The noise of this phase measurement is limited by the SQL to $\Delta\theta \geq N^{-1/2}$ for a given total number $N$ of unentangled particles.
The phase resolution can surpass the SQL by employing entangled particles, and reaches down to the ultimate Heisenberg limit $\Delta\theta \geq N^{-1}$.
A phase resolution near the Heisenberg limit requires the counting of atoms with single-particle resolution.
The need for single-particle resolution can only be avoided by the application of echo protocols, where a nonlinear squeezing interaction first reduces the quantum fluctuations before the sensing and then increases both signal and fluctuations before a -- possibly noisy -- detection~\cite{Davis2016,Hosten2016a,Anders2018,Schulte2019}.
However, such echo schemes can only be implemented in very specific measurement tasks.

Apart from entanglement-enhanced metrology, entangled many-particle states may also be exploited as a resource for quantum information.
Entanglement that is created between indistinguishable atoms in Bose-Einstein condensates (BECs) can be transferred to entanglement between spatially addressable subsystems~\cite{Lange2018,Kunkel2018,Fadel2018}.
Such systems allow violating a multi-particle Bell inequality~\cite{Laloee2009}, which could be a first step towards more advanced quantum information protocols.

One technique for measuring the number of neutral atoms in ultra-cold  ensembles with single-particle resolution is realised by loading the atoms into a magneto-optical trap (MOT)~\cite{Hume2013}.
The method has been extended to two separate spatial modes, as is required for an application in atom interferometry~\cite{Stroescu2015}.
These experiments demonstrated single-particle resolving atom counting for up to 1200 atoms.
In this sense, the atomic detection outperforms the capabilities in the analysis of indistinguishable photonic quantum states~\cite{Calkins2013,Harder2016}.
However, a single-atom resolving detection has still to be combined with an apparatus for the generation of entanglement in BECs.

In this article, we present a single-particle resolving atom number detection in a MOT which is fully integrated into an apparatus for the generation of many-particle entangled quantum states.
We demonstrate the counting of up to 30 atoms with single-particle resolution.
According to our noise analysis, we extrapolate that the single-atom resolution extends to a limiting atom number of $390(20)$ atoms.

Accurate detection techniques also aid the creation of desired atomic ensemble sizes with sub-Poissonian number fluctuations.
The high-fidelity preparation of a few-fermion system in its ground state was verified using fluorescence detection in a MOT configuration~\cite{Serwane2011}, while a non-destructive Faraday imaging technique has been utilized to prepare ultracold atom clouds at the shot noise level~\cite{Gajdacz2016,Kristensen2017}.
We apply the single-atom resolving detection to demonstrate a novel preparation of laser-cooled atomic samples.
In a feedback loop, a dedicated loss process steadily reduces the number of atoms, while the atom number of the ensemble is regularly measured, until a target number is reached.
We demonstrate the controlled preparation of $7$ atoms  with a fidelity of $92(2)\,\%$, which corresponds to a suppression of the number fluctuations by $18(1)\,\mathrm{dB}$ below the shot noise level.
We propose that the developed technique can also be employed to improve the counting capabilities under the influence of slow drifts.
In the future, the developed single-atom resolving detection can be utilized for obtaining unprecedented fidelities in the analysis of many-particle quantum states.

\section{Fluorescence imaging of individual atoms in a magneto-optical trap}
The detection of ultracold atomic ensembles is typically realized by a short illumination of the freely falling cloud with resonant light.
During the illumination of a few tens of microseconds, either the unabsorbed or the scattered photons are collected, until the atomic sample is accelerated and diluted too much due to the strong light pressure.
The obtainable resolution in the counting of atoms is ultimately limited by the shot noise of the detected photons, or even more precisely, by the shot noise of the photoelectrons that are counted in the detector.
Much longer detection times can be reached, when the atoms are trapped during illumination.
The additional trapping can be realized by far-detuned optical lattices~\cite{Sherson2010,Bakr2009,Eliasson2019}, optical dipole traps~\cite{Endres2016,Barredo2016,OhldeMello2019} or MOTs~\cite{Raab1987,Hu1994,Haubrich1996,Ruschewitz1996,Yoon2006,Hume2013}.
In such traps, lifetimes above one second can be reached which allow for a greatly improved signal-to-noise ratio.

\subsection{Emission and detection of photons}
In our experiments, the detection system is integrated into an apparatus that allows for the fast creation of many-particle entangled states in a BEC.
The systems includes a 3D-MOT with large $14\,\mathrm{mm}$ beams that is loaded by a 2D\textsuperscript{+}-MOT~\cite{Dieckmann1998} at a rate of $9.5(1)\times10^9$ atoms per second, coils for a magnetic quadrupole trap with a gradient of $300\,\mathrm{G}/\mathrm{cm}$, and a crossed-beam dipole trap with a wavelength of $1064\,\mathrm{nm}$.
For the detection, we trap $^{87}\mathrm{Rb}$ atoms in an additional MOT with small beams.
The atoms are imaged onto a charge-coupled-device (CCD) camera with a photon detection efficiency of $97\%$.
A large fraction ($5\,\%$) of the scattered photons are captured by a custom designed  objective with a high numerical aperture of $\mathrm{NA}=0.45$.
Despite its high  $\mathrm{NA}$, the objective can be placed outside the vacuum cell, due to the combination of a large diamater of $50\,\mathrm{mm}$ and a large working distance of $48\,\mathrm{mm}$ of the first aspheric lens.
A second plano-convex lens with a diameter of $75\,\mathrm{mm}$ and a focal length of $142\,\mathrm{mm}$ completes the objective, which is housed inside an aluminum tube.
The inside of the tube is painted with blackboard coating, reducing the transmission of unwanted background light.
With a total magnification of $M=2.62$ the resolution of the imaging system is limited by the pixel size of the CCD camera to about $5\,\mathrm{\mu m}$.
\begin{figure}
\includegraphics[scale=1]{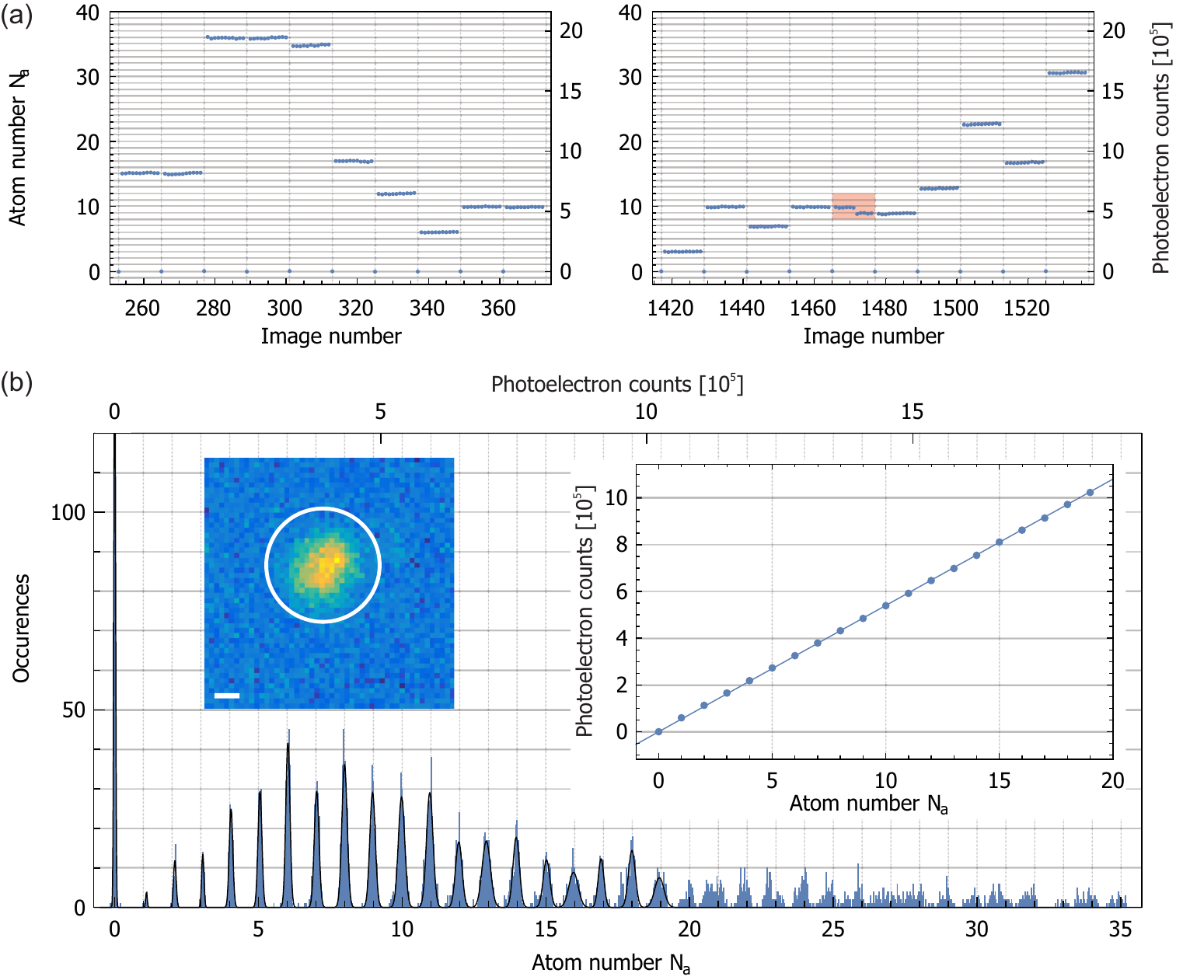}
  \caption{Example time traces, single atom image and fluorescence level histogram. (a) Two panels show example time traces of the recorded fluorescence signal in units of atom number (left axis) and photoelectrons (right axis). While the atom number is generally stable, in rare cases the loss of a single atom is recorded (red highlight). (b) The histogram of all recorded fluorescence images is fit with a sum of gaussian distributions. Inset left: Fluorescence image of a single atom trapped in the detection MOT. The scale bar measures $50\,\mathrm{\mu m}$. The white circle exemplifies the region of interest over which counts are integrated. Inset right: The center positions of the Gaussian fits show a linear dependence on the atom number and are used to extract the atom number to photoelectron calibration factor of $600(1)\times 10^3\,\mathrm{photoelectrons}/\mathrm{atom}/\mathrm{s}$.}
  \label{fig:SingleAtomFluorescence}
\end{figure}
For an evaluation of the detection system, the recorded images of the trapped atomic cloud are processed as follows.
The overall signal of a cloud of $N_\mathrm{a}$ atoms is determined by summing the counted photoelectrons over the area of the MOT image on the CCD camera (see figure \ref{fig:SingleAtomFluorescence}(b)).
Each atom within the MOT contributes a signal of $n_\mathrm{ph}=R_{\mathrm{sc}}\tau_\mathrm{det}\eta$ photoelectrons, where $R_{\mathrm{sc}}$ is the photon scattering rate, $\tau_\mathrm{det}$ the detection time and $\eta$ the overall detection efficiency.
The photon scattering rate is given by $R_\mathrm{sc}=\Gamma/2 \times s_0 /\left(1+s_0+4\Delta^2/\Gamma^2\right)$, where $\Gamma=2\pi\times6\,\mathrm{MHz}$ is the natural line width of the $^{87}\mathrm{Rb}$ $\mathrm{D}_2$ transition, $\Delta$ represents the detuning of the laser with respect to the resonance frequency of the transition and $s_0=I/I_\mathrm{sat}$ is the saturation parameter that describes the ratio of the collective intensity $I$ of the laser beams and the isotropic saturation intensity $I_\mathrm{sat}=3.576\,\mathrm{mW}/\mathrm{cm}^2$.
The photoelectron shot noise contributes a term of $\sigma^2_\mathrm{psn}=N_\mathrm{a}/n_\mathrm{ph}$ to the total signal variance and can thus be reduced by extending the exposure time.
The contrary holds true for the noise caused by atom loss.
Due to the finite lifetime of the trap, atom loss contributes a term $\sigma^2_\mathrm{loss}=N_\mathrm{a}\tau_\mathrm{det}/\tau_\mathrm{life}$, where $\tau_\mathrm{life}$ is the lifetime of the MOT.
By employing small detection MOT beams with a diameter of $w=1.25\,\mathrm{mm}$, potential stray light sources causing unwanted background noise in the images are reduced.
With a collective peak intensity of $24\,\mathrm{mW/cm}^2$, the three beam pairs yield a combined saturation parameter of $s_0=6.65$.
Together with the detuning of $\Delta=2\pi\times 6\,\mathrm{MHz}$, the scattering rate is estimated to be $R_\mathrm{sc}=1.1\times10^7\,\mathrm{photons/s}$.
During an exposure of $90\,\mathrm{ms}$, our detection system with its total efficiency of $\eta=4.71\,\%$ is expected to collect $4.7\times10^4$ photoelectrons per atom.

\subsection{Identification of single atoms}
Our experimental procedure for the calibration of the atom number detection starts out by the acquisition of a background image without atoms but with all relevant light sources.
This is followed by the loading of atoms from the 2D\textsuperscript{+}-MOT into the 3D-MOT configuration for a duration of only $15\,\mathrm{ms}$.
Afterwards the trapping light is switched off for a short duration of $10\,\mathrm{ms}$ before a small number of atoms is recaptured from the expanding cloud by activating our small detection MOT with  $1.25\,\mathrm{mm}$-diameter beams.
Here, the atomic ensemble is held for $500\,\mathrm{ms}$ in order to ensure that the remaining untrapped atoms have left the detection region.
Now, the fluorescence signal of the atomic sample is acquired over a time of $\tau_\mathrm{det}=90\,\mathrm{ms}$.
During camera read-out, the atom cloud is held in the trap for $220\,\mathrm{ms}$ before another image is taken with the same exposure time.
These two steps are repeated until a total set of 11 images is acquired.
The time traces of the fluorescence signal (see figure \ref{fig:SingleAtomFluorescence}(a)) exhibit clearly discernable levels.
In rare cases, unwanted single-atom loss or loading events can be observed.
A histogram based on more than $5200$ recorded atom images shows well-resolved peaks for up to $N_\mathrm{a}\approx 30$ atoms.
The peaks reflect the integer number of atoms that are held in the trap.
The clear visibility of these features proves that the resolution is well below the single-atom level.
Fitting a sum of Gaussian functions to the first 20 peaks of the histogram reveals the center positions of the individual peaks.
These center positions scale linearly with the detected camera counts, corresponding to single-atom count rate of $600(1)\times10^3$ photoelectron counts per atom per second (see inset in figure \ref{fig:SingleAtomFluorescence}(b)), which matches our expectation to within $10\%$. 
This calibration yields an accurate absolute value for the number of atoms without the need of a precise specification of the experimental parameters such as laser powers, laser detunings and beam sizes.

\section{Lifetime and loading rate analysis}
\begin{figure}
\includegraphics[scale=1]{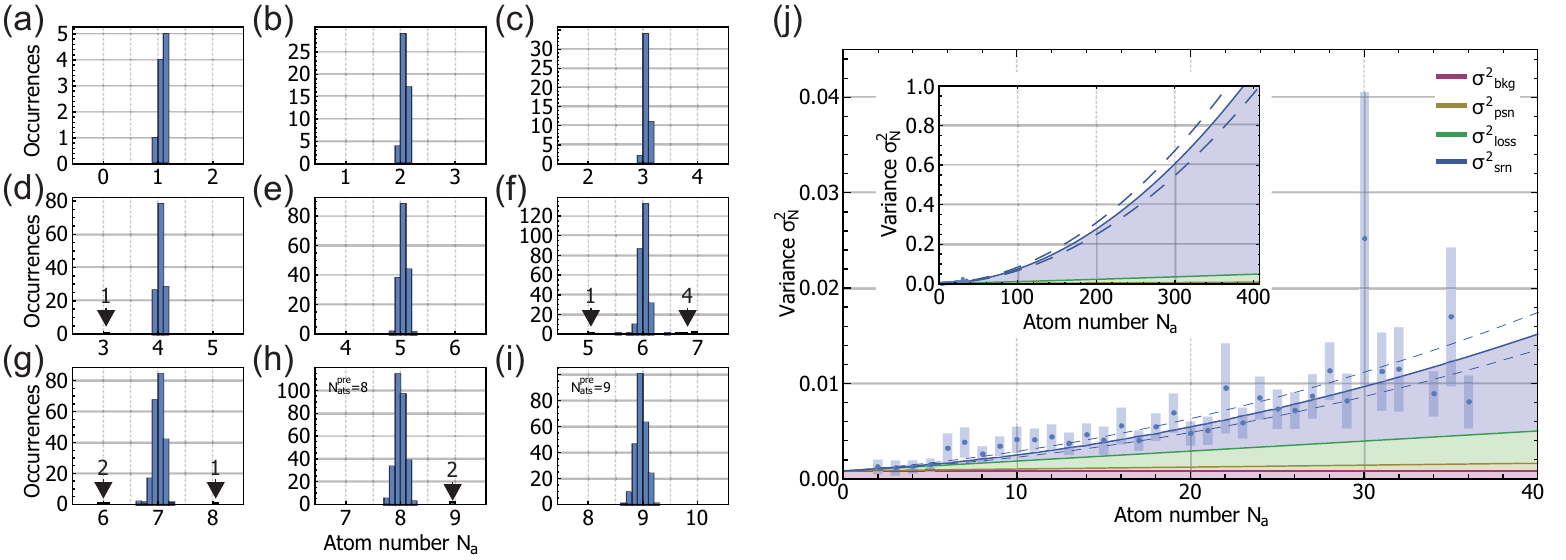}
  \caption{Analysis of consecutive fluorescence levels and noise model fit. (a-i) By comparing consecutively recorded fluorescence levels single atom loading and loss events are extracted. The histograms exemplify the event classification depending on the initial atom number from $N_{\mathrm{a}}=1$ to $N_{\mathrm{a}}=9$, respectively. The central peak in each histogram represents the survival of all atoms present in the MOT, while occurrences to its left and right correspond to loss and loading events, respectively. (j) The noise model is fit to the experimental data. The colors indicate the contributions from different noise sources. Inset: An extrapolation of the noise model fit suggests an upper limit of $390(20)$ atoms at which individual atoms become indistinguishable by our current detection setup.}
\label{fig:StepStatisticsAndNoiseModel}
\end{figure}

For optimal performance of the detection setup, a long lifetime of the MOT is crucial as it limits the usable illumination time.
Similarly, the detection benefits from a small loading rate, caused by atoms being captured from the background gas or the residual flow from the 2D\textsuperscript{+}-MOT cell.
Both parameters, lifetime and loading rate, can be extracted from the recorded time traces.
We evaluate each possible pair of successive measurements and classify them as loss, loading or survival event for atom numbers up to $N_\mathrm{a}=15$.
For each atom number between $N_\mathrm{a}=1$ and $N_\mathrm{a}=9$, a histogram showcases the occurrences of those three events in figure \ref{fig:StepStatisticsAndNoiseModel}(a-i).
Importantly, across the full data set no two-body loss events were identified.
This fact in conjunction with the occurrence of only 14 loss events across 24482 observed individual atoms in the image pairs shows, that the total holding and detection time of $\tau_\mathrm{hold}+\tau_\mathrm{det}=310\,\mathrm{ms}$ is short compared to the lifetime of the trap.
The loss process can be expected to follow Poissonian statistics, such that the lifetime is $\tau_\mathrm{life}=\tau_\mathrm{hold}+\tau_\mathrm{det}/P_\mathrm{loss}\left(\tau_\mathrm{hold}+\tau_\mathrm{det}\right)=540(140)\,\mathrm{s}$, where $P_\mathrm{loss}\left(\Delta t\right)$ is the probability for a loss event to occur during the time span $\Delta t$.
The loading rate $R_\mathrm{load}=0.014(4)\,\mathrm{s}^{-1}$ is based on 12 observed events within a set of 2710 image pairs and is a result of the low capture velocity in combination with a low Rb background pressure.
In total, these measurements prove that the single-atom resolution will not be limited by finite lifetime or residual loading, even for an improvement towards larger ensembles.

\section{Noise model}
The capability of our number detection is best analysed based on the shot-to-shot number counting fluctuations, where long-term drifts of the scattering rate, which may be caused by intensity or frequency drifts in the laser light, are not considered.
These shot-to-shot fluctuations can be described by the two-sample variance $\sigma^2_N=1/2\langle\left(N_{\mathrm{a},j+1}-N_{\mathrm{a},j}\right)^2\rangle_j$, where $j$ is an index running across successively captured images.
Our noise model
\begin{equation}
\sigma^2_N=\sigma^2_\mathrm{bkg}+\sigma^2_\mathrm{psn}+\sigma^2_\mathrm{srn}+\sigma^2_\mathrm{loss}
\end{equation}
includes contributions from background noise, photoelectron shot noise, scattering rate noise and noise from atom loss.
From the acquired background images, the background contribution is calculated to be $\sigma^2_\mathrm{bkg}=8.4\times10^{-4}$ and hence well below the single atom level.
The photoelectron shot noise $\sigma^2_\mathrm{psn}=N_\mathrm{a}/\left(\eta\tau_\mathrm{det}R_\mathrm{sc}\right)$ scales linearly with the atom number.
The scattering rate noise $\sigma^2_\mathrm{srn}=N^2_\mathrm{a}\alpha^2/\tau_ \mathrm{det}$ is caused by corresponding fluctuations in the intensity and frequency of the MOT light that are combined into the fluorescence noise parameter $\alpha$.
Finally we consider the linear single-atom loss term $\sigma^2_\mathrm{loss,lin}=\tau_\mathrm{det}N_\mathrm{a}/\left(2\tau_\mathrm{life}\right)$, while we found contributions from mean atom-loss and two-body losses due to light-assisted collisions to be negligible for our data \cite{Hume2013}.
The resulting noise model reads 
\begin{equation}
\sigma^2_N=\sigma^2_\mathrm{bkg}+N_\mathrm{a}\left(1/\left(\eta\tau_\mathrm{det}R_\mathrm{sc}\right)+\tau_\mathrm{det}/\left(2\tau_\mathrm{life}\right)\right)+N_\mathrm{a}^2\alpha^2/\tau_\mathrm{det}.
\end{equation}

By fitting the noise model to our data for atom numbers up to $N_\mathrm{a}=36$, with the fluorescence noise parameter as the only free parameter, we obtain a value of $\alpha=7.6(4)\times10^{-4}\,\mathrm{s}^{1/2}$.
This corresponds to either $22\,\mathrm{kHz}$ in laser frequency noise or to a relative fluctuation of $0.039$ in the saturation parameter $s_0$.
Extrapolating the noise model to the critical single-atom detection threshold $\sigma^2_N=1$ yields a maximally discernable atom number of $N_\mathrm{a}^{\mathrm{max}}=390(20)$ atoms.

\section{Atom-number stabilization}
\begin{figure}
\includegraphics[scale=1]{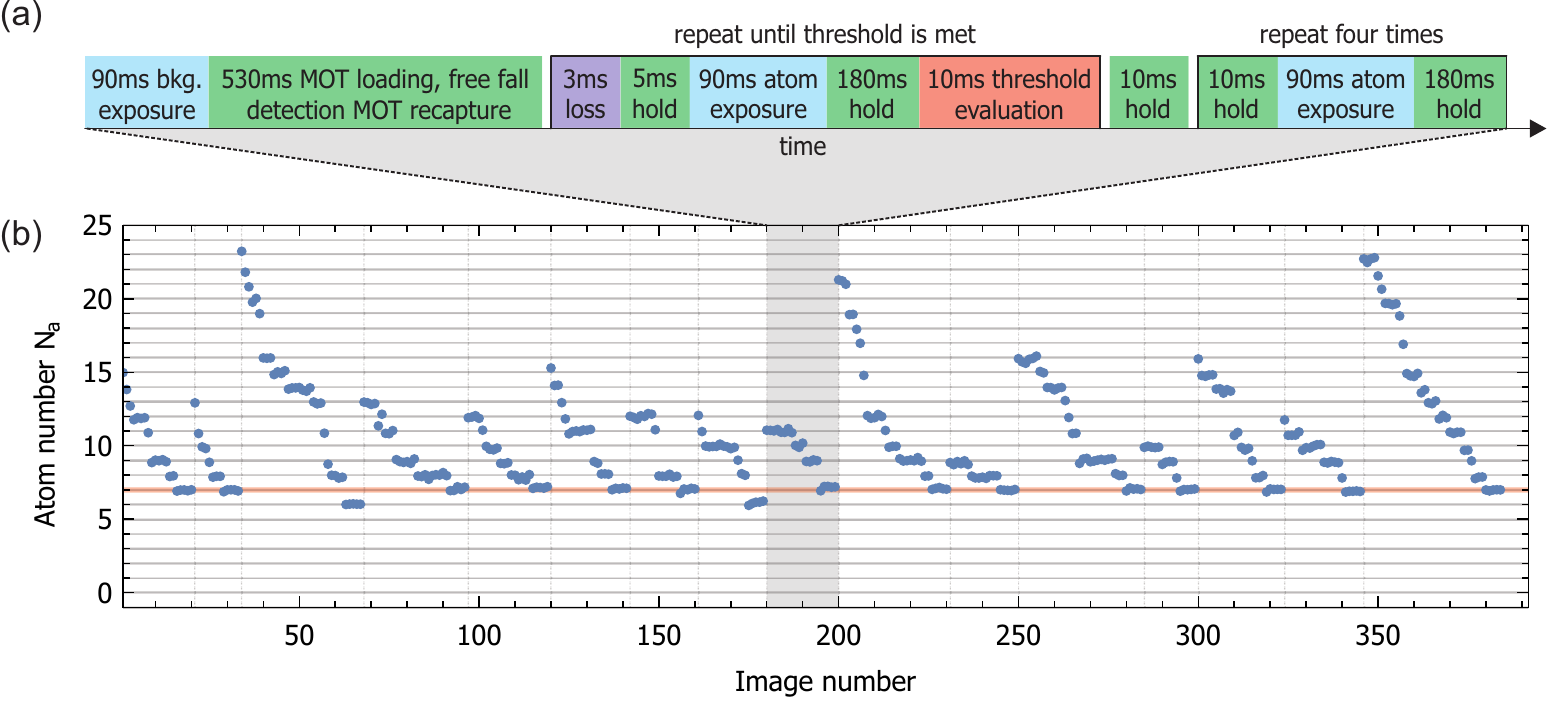}
  \caption{(a) Experimental sequence for the atom number stabilization. (b) 15 example time traces ending in stabilized atom numbers with varying initial atom numbers in the MOT. A new measurement is started (vertical gray lines) once the target atom number of $7$ (red horizontal line) has been hit or undercut for five successive images.}
\label{fig:StabilizationTimeTraces}
\end{figure}
\begin{figure}
\includegraphics[scale=1]{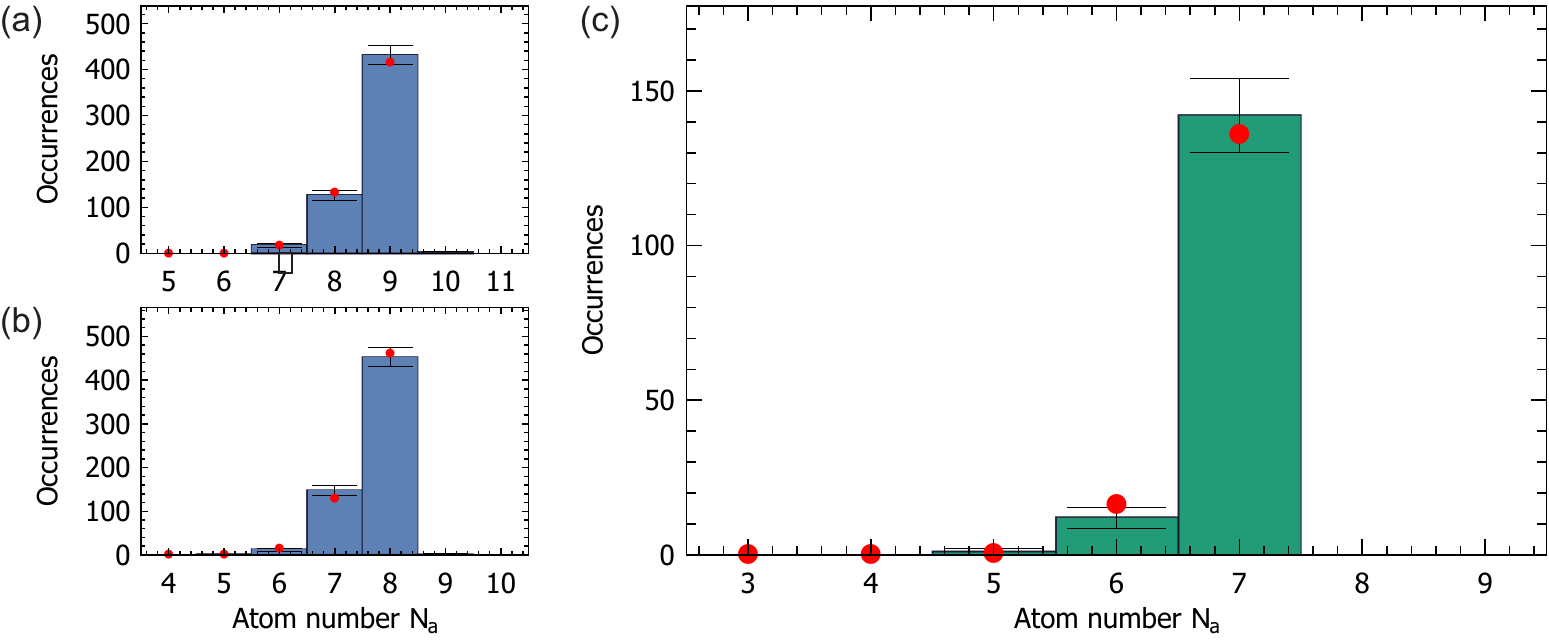}
  \caption{Loss characterization and stabilization result. (a,b) The exemplary histograms show the transition statistics of consecutive measurements for initial atom numbers $N_\mathrm{a}^\mathrm{(i)}=9$ and $N_\mathrm{a}^\mathrm{(i)}=8$ with the corresponding statistical errors. A collectively fit binomial model (red points) for the transitions statistics from $N_\mathrm{a}^\mathrm{(i)}=18$ to $N_\mathrm{a}^\mathrm{(i)}=8$ results in a survival probability of $p_\mathrm{s}=96.66(4)\,\%$ for an individual atom over the course of a single loss step. (c) The histogram shows the atom number distribution for the successful stabilization with the corresponding statistical errors. It is derived from the atom number evaluation of the threshold image once the loss process is stopped. An ensemble of $7$ atoms was prepared with a fidelity of $92(2)\,\%$, corresponding to an $18(1)\,\mathrm{dB}$ suppression below shot noise. A truncated binomial distribution for our derived survival probability centered around $8$ atoms describes the data accurately (red points).}
\label{fig:StabilizationStepStatistics}
\end{figure}
We utilize our accurate atom counting to deliver precise atom numbers on demand, by interleaving the atom number measurement with a dedicated loss process, until a desired number is reached.
We start out with an average cloud of $\langle N_\mathrm{a} \rangle=15(4)$ atoms trapped in our detection MOT.
Upon image acquisition, the atom number is estimated in real time.
A field-programmable gate array (FPGA), that controls the experimental protocol stops the loss sequence once the detected number of atoms falls below a desired threshold of $N_\mathrm{a}^\mathrm{thr}=7.5$ atoms. The prepared ensemble is stored in the MOT to check the final atom number with four additional number measurements.
The loss is induced by turning off the repumping light for $3\,\mathrm{ms}$ during the $198\,\mathrm{ms}$ of holding time between the $90\,\mathrm{ms}$ detection windows (see figure \ref{fig:StabilizationTimeTraces}(a)).
Time traces in figure \ref{fig:StabilizationTimeTraces}(b) show that the fluorescence level of the MOT detection halts at our desired atom number of $N_\mathrm{a}=7$.
All atoms have an independent and equal survival probability $p_\mathrm{s}$ with which they remain in the trap.
Each loss step can be viewed as an independent series of Bernoulli trials, such that the atom number statistics will follow a binomial distribution.
The histograms in figure \ref{fig:StabilizationStepStatistics}(a,b) showcase the transitions from an input atom number $N_\mathrm{a}^\mathrm{(i)}$ to an output atom number $N_\mathrm{a}^\mathrm{(s)}$ for a single loss step.
Collectively fitting the histogram data with a binomial distribution
\begin{equation}
B\left(p_\mathrm{s},N_\mathrm{a}^\mathrm{(i)},N_\mathrm{a}^\mathrm{(s)}\right)=\binom{N_\mathrm{a}^\mathrm{(i)}}{N_\mathrm{a}^\mathrm{(s)}}p_\mathrm{s}^{N_\mathrm{a}^\mathrm{(s)}}\left(1-p_\mathrm{s}\right)^{N_\mathrm{a}^\mathrm{(i)}-N_\mathrm{a}^\mathrm{(s)}}
\end{equation}
reveals their common survival probability $p_\mathrm{s}=96.66(4)\,\%$, that characterizes our loss process.
For a given survival probability, it is possible to calculate a maximal obtainable preparation fidelity because of the unwanted accidental removal of two atoms in the final loss step.
For our loss step, we obtain a fidelity of $88\,\%$ to obtain a state with exactly $7$ atoms.
This corresponds to a suppression of the number fluctuations of  $17.3\,\mathrm{dB}$ below shot noise.
Due to the high survival probability, single-step jumps from higher atom numbers (the dominant unwanted process would be a jump from 9 to 6 atoms) do not contribute to the obtained results.
Figure~\ref{fig:StabilizationStepStatistics}(c) shows the final result.
We obtain the target number of $7$ atoms in  $142$ of the $155$ cases, which corresponds to a state preparation with a $92(2)\,\%$ fidelity.
We obtain a too small result in $13$ cases ($6$ atoms: $12$ cases, $5$ atoms: $1$ case).
From these data, we extract a suppression of $18(1)\,\mathrm{dB}$.

\section{Summary}
In summary, we have demonstrated a single-atom resolving detection of up to $30$ atoms in an experimental apparatus that is designed for the generation of entangled many-particle states.
The shot-to-shot fluctuations suggest that the scattering rate noise limits the single-atom resolution to $390(20)$ atoms.
In conjunction with our accurate atom number detection, we have employed a real-time feedback onto the repumping light of the MOT to stabilize the number of atoms in the laser-cooled ensemble.
A preparation fidelity of $92(2)\,\%$ was demonstrated for an ensemble of $7$ atoms, corresponding to a suppression of $18(1)\,\mathrm{dB}$ below the shot noise level.
This technique allows to deliver number-stabilizied atomic ensembles on demand.

Interestingly, the loss procedure can be employed to overcome the influence of slow drifts in the scattering rate.
If such drifts deteriorate the detection of a large number of several hundreds atoms, the initial illumination can be followed by several iterations of engineered loss and detection.
Thereby, one obtains a series of number measurements that optimally spans the full range of atom numbers between the initial number and zero.
This series of number measurements allows for an individual calibration of the current scattering rate, as each individual number measurement must correspond to an integer number.

In the future, we will apply the developed  detection to analyze many-particle quantum states with single-particle resolution as well as advance our methods for metrology beyond the SQL~\cite{Lucke2011,Kruse2016} towards the ultimate Heisenberg limit.

We acknowledge support from the Deutsche Forschungsgemeinschaft (DFG) through CRC 1227 (DQ-mat), project B01.

\section*{References}
\bibliography{main}

\providecommand{\newblock}{}
\begin{thebibliography}{10}
\expandafter\ifx\csname url\endcsname\relax
  \def\url#1{{\tt #1}}\fi
\expandafter\ifx\csname urlprefix\endcsname\relax\def\urlprefix{URL }\fi
\providecommand{\eprint}[2][]{\url{#2}}

\bibitem{McConnell2015}
McConnell R, Zhang H, Hu J, \'Cuk S and Vuleti\'c V 2015 {\em Nature\/} {\bf
  519} 439--442

\bibitem{Haas2014}
Haas F, Volz J, Gehr R, Reichel J and Est{\`e}ve J 2014 {\em Science\/} {\bf
  344} 180--183

\bibitem{Lucke2014}
L\"ucke B, Peise J, Vitagliano G, Arlt J, Santos L, T\'oth G and Klempt C 2014
  {\em Phys. Rev. Lett.\/} {\bf 112}(15) 155304

\bibitem{Hosten2016}
Hosten O, Engelsen N~J, Krishnakumar R and Kasevich M~A 2016 {\em Nature\/}
  {\bf 529} 505

\bibitem{Pezze2018}
Pezz{\`{e}} L, Smerzi A, Oberthaler M~K, Schmied R and Treutlein P 2018 {\em
  Rev. Mod. Phys.\/} {\bf 90}

\bibitem{Davis2016}
Davis E, Bentsen G and Schleier-Smith M 2016 {\em Phys. Rev. Lett.\/} {\bf
  116}(5) 053601

\bibitem{Hosten2016a}
Hosten O, Krishnakumar R, Engelsen N and Kasevich M 2016 {\em Science\/} {\bf
  352} 1552--1555

\bibitem{Anders2018}
Anders F, Pezz\`e L, Smerzi A and Klempt C 2018 {\em Phys. Rev. A\/} {\bf
  97}(4) 043813

\bibitem{Schulte2019}
Schulte M, Mart{\'i}nez-Lahuerta V~J, Scharnagl M~S and Hammerer K {\em
  arXiv:1911.11801v1\/}

\bibitem{Lange2018}
Lange K, Peise J, L{\"u}cke B, Kruse I, Vitagliano G, Apellaniz I, Kleinmann M,
  T\'oth G and Klempt C 2018 {\em Science\/} {\bf 360} 416--418

\bibitem{Kunkel2018}
Kunkel P, Pr{\"u}fer M, Strobel H, Linnemann D, Fr{\"o}lian A, Gasenzer T,
  G{\"a}rttner M and Oberthaler M~K 2018 {\em Science\/} {\bf 360} 413--416

\bibitem{Fadel2018}
Fadel M, Zibold T, D{\'e}camps B and Treutlein P 2018 {\em Science\/} {\bf 360}
  409--413

\bibitem{Laloee2009}
Lalo{\"e} F and Mullin W~J 2009 {\em Eur. Phys. J. B\/} {\bf 70} 377--396

\bibitem{Hume2013}
Hume D~B, Stroescu I, Joos M, Muessel W, Strobel H and Oberthaler M~K 2013 {\em
  Phys. Rev. Lett.\/} {\bf 111}(25) 253001

\bibitem{Stroescu2015}
Stroescu I, Hume D~B and Oberthaler M~K 2015 {\em Phys. Rev. A\/} {\bf 91}
  013412

\bibitem{Calkins2013}
Calkins B, Mennea P~L, Lita A~E, Metcalf B~J, Kolthammer W~S, Lamas-Linares A,
  Spring J~B, Humphreys P~C, Mirin R~P, Gates J~C, Smith P~G~R, Walmsley I~A,
  Gerrits T and Nam S~W 2013 {\em Opt. Express\/} {\bf 21} 22657

\bibitem{Harder2016}
Harder G, Bartley T~J, Lita A~E, Nam S~W, Gerrits T and Silberhorn C 2016 {\em
  Phys. Rev. Lett.\/} {\bf 116}

\bibitem{Serwane2011}
Serwane F, Z{\"u}rn G, Lompe T, Ottenstein T~B, Wenz A~N and Jochim S 2011 {\em
  Science\/} {\bf 332} 336--338

\bibitem{Gajdacz2016}
Gajdacz M, Hilliard A~J, Kristensen M~A, Pedersen P~L, Klempt C, Arlt J~J and
  Sherson J~F 2016 {\em Phys. Rev. Lett.\/} {\bf 117}(7) 073604

\bibitem{Kristensen2017}
Kristensen M~A, Gajdacz M, Pedersen P~L, Klempt C, Sherson J~F, Arlt J~J and
  Hilliard A~J 2017 {\em J. Phys. B: At., Mol. Opt. Phys.\/} {\bf 50} 034004

\bibitem{Sherson2010}
Sherson J~F, Weitenberg C, Endres M, Cheneau M, Bloch I and Kuhr S 2010 {\em
  Nature\/} {\bf 467} 68--72

\bibitem{Bakr2009}
Bakr W~S, Gillen J~I, Peng A, Folling S and Greiner M 2009 {\em Nature\/} {\bf
  462} 74--77

\bibitem{Eliasson2019}
El\'iasson O, Heck R, Laustsen J~S, M\"uller R, Weidner C~A, Arlt J~J and
  Sherson J~F 2019 {\em arXiv:1912.03079\/}

\bibitem{Endres2016}
Endres M, Bernien H, Keesling A, Levine H, Anschuetz E~R, Krajenbrink A, Senko
  C, Vuletic V, Greiner M and Lukin M~D 2016 {\em Science\/} {\bf 354}
  1024--1027

\bibitem{Barredo2016}
Barredo D, de~L{\'e}s{\'e}leuc S, Lienhard V, Lahaye T and Browaeys A 2016 {\em
  Science\/} {\bf 354} 1021--1023

\bibitem{OhldeMello2019}
Ohl~de Mello D, Sch\"affner D, Werkmann J, Preuschoff T, Kohfahl L, Schlosser M
  and Birkl G 2019 {\em Phys. Rev. Lett.\/} {\bf 122}(20) 203601

\bibitem{Raab1987}
Raab E~L, Prentiss M, Cable A, Chu S and Pritchard D~E 1987 {\em Phys. Rev.
  Lett.\/} {\bf 59} 2631--2634

\bibitem{Hu1994}
Hu Z and Kimble H~J 1994 {\em Opt. Lett.\/} {\bf 19} 1888

\bibitem{Haubrich1996}
Haubrich D, Schadwinkel H, Strauch F, Ueberholz B, Wynands R and Meschede D
  1996 {\em Europhys. Lett.\/} {\bf 34} 663--668

\bibitem{Ruschewitz1996}
Ruschewitz F, Bettermann D, Peng J~L and Ertmer W 1996 {\em Europhys. Lett.\/}
  {\bf 34} 651--656

\bibitem{Yoon2006}
Yoon S, Choi Y, Park S, Kim J, Lee J~H and An K 2006 {\em Appl. Phys. Lett.\/}
  {\bf 88} 211104

\bibitem{Dieckmann1998}
Dieckmann K, Spreeuw R~J~C, Weidem\"uller M and Walraven J~T~M 1998 {\em Phys.
  Rev. A\/} {\bf 58} 3891--3895

\bibitem{Lucke2011}
L{\"u}cke B, Scherer M, Kruse J, Pezz{\'e} L, Deuretzbacher F, Hyllus P, Topic
  O, Peise J, Ertmer W, Arlt J, Santos L, Smerzi A and Klempt C 2011 {\em
  Science\/} {\bf 334} 773--776

\bibitem{Kruse2016}
Kruse I, Lange K, Peise J, L\"ucke B, Pezz\`e L, Arlt J, Ertmer W, Lisdat C,
  Santos L, Smerzi A and Klempt C 2016 {\em Phys. Rev. Lett.\/} {\bf 117}(14)
  143004

\end{thebibliography}
\bibliographystyle{iopart-num}

\end{document}